\documentclass[runningheads]{llncs}
\usepackage{graphicx}
\usepackage[caption=false]{subfig}
\usepackage{booktabs}
\usepackage[sort]{cite}
\usepackage{listings}
\usepackage{amssymb}
\usepackage{multirow}
\usepackage{graphicx}
\usepackage{tabularx}
\usepackage{mathtools}
\newcolumntype{Y}{>{\centering\arraybackslash}X}
\usepackage[breaklinks=true,colorlinks,bookmarks=false]{hyperref}

\begin{document}
\title{EmbNum: Semantic Labeling for Numerical Values with Deep Metric Learning}
\titlerunning{EmbNum}

\author{
Phuc Nguyen\inst{1,2} \and
Khai Nguyen\inst{2}\and
Ryutaro Ichise\inst{1,2}\and
Hideaki Takeda\inst{1,2}
}

\institute
{
SOKENDAI (The Graduate University for Advanced Studies) \\ 
Shonan Village, Hayama, Kanagawa, Japan \and 
National Institute of Informatics \\ 
2-1-2 Hitotsubashi, Chiyoda-ku, Tokyo, Japan \\
\email{\{phucnt, nhkhai, ichise, takeda\}@nii.ac.jp}
}
\maketitle

\begin{abstract}
Semantic labeling for numerical values is a task of assigning semantic labels to unknown numerical attributes. The semantic labels could be numerical properties in ontologies, instances in knowledge bases, or labeled data that are manually annotated by domain experts. In this paper, we refer to semantic labeling as a retrieval setting where the label of an unknown attribute is assigned by the label of the most relevant attribute in labeled data. One of the greatest challenges is that an unknown attribute rarely has the same set of values with the similar one in the labeled data. To overcome the issue, statistical interpretation of value distribution is taken into account. However, the existing studies assume a specific form of distribution. It is not appropriate in particular to apply open data where there is no knowledge of data in advance. To address these problems, we propose a neural numerical embedding model (\textit{EmbNum}) to learn useful representation vectors for numerical attributes without prior assumptions on the distribution of data. Then, the ``semantic similarities" between the attributes are measured on these representation vectors by the Euclidean distance. Our empirical experiments on City Data and Open Data show that \textit{EmbNum} significantly outperforms state-of-the-art methods for the task of numerical attribute semantic labeling regarding effectiveness and efficiency.
\end{abstract}

\section{Introduction}
Thanks to the Open Data movement, a large number of table data resources have been published on the Web or Open Data portals. In a study of Lehmberg et al., 233 million tables were extracted from the July 2015 version of the Common Crawl\footnote{http://commoncrawl.org/} \cite{Lehmberg2016}. Additionally, Mitlohner et al. performed characteristics analysis of 200,000 tables from 232 Open Data portals \cite{Mitlohner2016}. These resources could be integrated, and enabling them to be potentially useful for other applications such as table search \cite{Nguyen2015, Nargesian2018}, table extension \cite{Lehmberg2015}, completion\cite{ahmadov2015towards}, or knowledge base construction as used in DBpedia \cite{Zhang2013}, YAGO \cite{Sekhavat2014}, Freebase \cite{Dong2014}. 

However, these data resources are very heterogeneous. Each data resource is independently constructed by different people with different backgrounds, purposes, and contexts. Therefore, the use of vocabulary and schema structure might differ across various data resources. For example, one attribute uses ``population" as the table header label; another table uses ``number of people." Do those two attributes labels share the same meaning or different ones? As a result, the ``semantic heterogeneity" may lead to the propagation of misinformation in the data integrating process.

To provide a unified view of the heterogeneous resources, one of the possible solutions is to assign a semantic label for each attribute in unknown resources. The semantic labels could be properties in ontologies, instances in knowledge bases, or labeled attributes manually annotated by domain experts. In this paper, the problem of semantic labeling is formulated as a retrieval setting where the label of an unknown attribute is assigned by the label of the most relevant attribute in labeled data. In other words, given a query as a list of values of an unknown attribute, the system will return a ranking list of the relevant labeled attributes with respect to a specific similarity metric. The label of the unknown attribute is assigned by the label of the most relevant one in the ranking list.

The most common approaches for semantic labeling use textual information, such as header labels, textual values, or table description. Previous studies \cite{Ermilov2013, Zhang2013, ritze2015matching, Adelfio2013, Venetis2011, Wang2012} used text-based entity linkage to search for similar concepts in knowledge bases. Then, semantic labels can be inferred by using semantic labels of matched entities in the knowledge base. However, many attributes do not have overlapping entity labels with the knowledge bases. Even if some data overlap, many attributes do not have similar labels with the entities in knowledge bases because these are expressed as numbers, IDs, codes, or abbreviations \cite{neumaier2016multi}. Therefore, a text-based approach cannot directly be used on these table data.

Another direction of semantic labeling uses numerical information. As mentioned in a study by Mitlohner et al., 50 \% of the table data taken from Open Data portals contain numerical values \cite{Mitlohner2016}. Prior studies \cite{neumaier2016multi, Stonebraker2013, ramnandan2015assigning, pham16:iswc} used descriptive statistics with hypothesis tests as a metric to compare the similarity of numerical attributes. However, these hypothesis tests often rely on the assumption that these attributes have to be drawn from a specific form of distribution (e.g., normal distribution or uniform distribution) or data types (e.g., continuous or discrete). Knowing the form of distributions and data types of unknown numerical attributes is a difficult challenge. As a result, a proper hypothesis test cannot be easily selected when we do not know data distribution and data type. 

In recent years, there has been an increasing interest in using deep metric learning to learn similarity metric directly from data \cite{sharif2014cnn, schroff2015facenet, hermans2017defense}. The principal advantage of this technique is an ability to learn the feature representation and the similarity metric in an end-to-end fashion. Inspired by their success and an assumption with shared meanings of numerical attributes across different data resources, we explored whether or not deep metric learning can be used to learn a metric in measuring the ``semantic similarity" of numerical attributes without making any assumption on data type, or data distribution. Indeed, we used a representation network consisting of a triplet network and convolutional neural network (CNN) to learn a non-linear mapping function from numerical attributes to a transformed space. In other words, the non-linear mapping function was used as an embedding model to derive the latent features for numerical attributes. The ``semantic similarity" of two numerical attributes was calculated on two extracted features using the Euclidean distance. 

We also introduced an inverse transform sampling \cite{wikiits} to deal with the issues of varying input size of table numerical attributes. The representation network required a fixed size as the input, but the size of numerical attributes could vary from a few numerical values to thousands of them \cite{Mitlohner2016}. The sampling technique can capture the original distribution of numerical attributes, thereby overcoming the issue of varying the size of the input attributes. Moreover, the sampling technique also helps to speed up data processing since a small number of data values is considered instead of entire values of the attribute. 

Overall, our contributions to this paper are as follows.
\begin{enumerate}
\item We propose a novel model called \textit{EmbNum} to learn the representation and the similarity metric from numerical attributes. \textit{EmbNum} is constructed by jointly learning the representation and the similarity metric from numerical attributes in an end-to-end setting. 
  \item We introduced an inverse transform sampling approach to handle the issue of varying the size of numerical attributes. The sampling technique can simulate the original distribution of numerical attributes.
  \item We created a new dataset (Open Data) extracted from tables of Open Data portals to test semantic labeling in the open environment. The dataset is available at https://github.com/phucty/embnum. 
  \item We conducted benchmarks of \textit{EmbNum} and two baseline approaches, i.e., SemanticTyper \cite{ramnandan2015assigning} and DSL \cite{pham16:iswc} on the standard data, e.g., City Data \cite{ramnandan2015assigning} and real-world data, e.g., Open Data. The overall results show that using \textit{EmbNum} achieved better performance of semantic labeling for numerical attributes on effectiveness and efficiency.
\end{enumerate}

The rest of this paper is organized as follows. In Section \ref{approach}, we present the preliminaries and overall framework of our approach. Section \ref{experiment} presents the details of our evaluation. In Section \ref{relate_work}, we discuss the related works in the task of semantic labeling for table data as well as representation learning techniques. Finally, we summarize the paper and discuss the future direction in Section \ref{conclusion}.

\section{Approach} \label{approach}
In this sections, we first describe the preliminary concepts (Section \ref{preli}) as well as the formal definition for semantic labeling. Then, we present the overall framework in Section \ref{overall}. The details explanation for each module of the framework is provided in Section \ref{preprocessing}, \ref{learning}, and \ref{labeling}.

\subsection{Preliminaries} \label{preli}
\subsubsection{Numerical attributes}
We call table columns as the attributes of tables. We assume that all values in one attribute have the same meaning and that shared meanings exist across tables. If all values in an attribute are numerical, we call the attribute as a numerical attribute. If not, we call it as a textual attribute. In this paper, we only consider a similarity metric between numerical attributes to infer semantic meaning. The similarity metric for a textual attribute is out of the scope of this paper. Notably, we did not tackle the problem of data scaling. For instance, if two attributes were found to have the same meaning, but they were expressed in different scales, we considered that the two attributes have a different meaning. Techniques for interpreting data scaling is left as future work.
\subsubsection{Semantic Labeling Definition}
Let $A =  \{ a_1, a_2, a_3, ..., a_n\}$ be a list of $n$ numerical attributes and $Y = \{ y_1, y_2, y_3, ..., y_m\}$ be a list of $m$ semantic labels, where $m \leqslant n$. We have a data sample $(a,y)$ with $a \in A$ and $y \in Y$ that is a pair of a numerical attribute and its semantic label. All data samples are stored in a $D$ database. 

Given an unknown attribute $a_q$, similarity searching is performed on all data samples in the $D$ database with respect to specific similarity metric. The output is a ranked list of the most relevant samples. The semantic label of the top result is assigned to the label of the unknown attribute. 

\subsection{Overall Approach} \label{overall}
\begin{figure}
\centering
\vspace*{-5mm}
\includegraphics[width=0.9\textwidth]{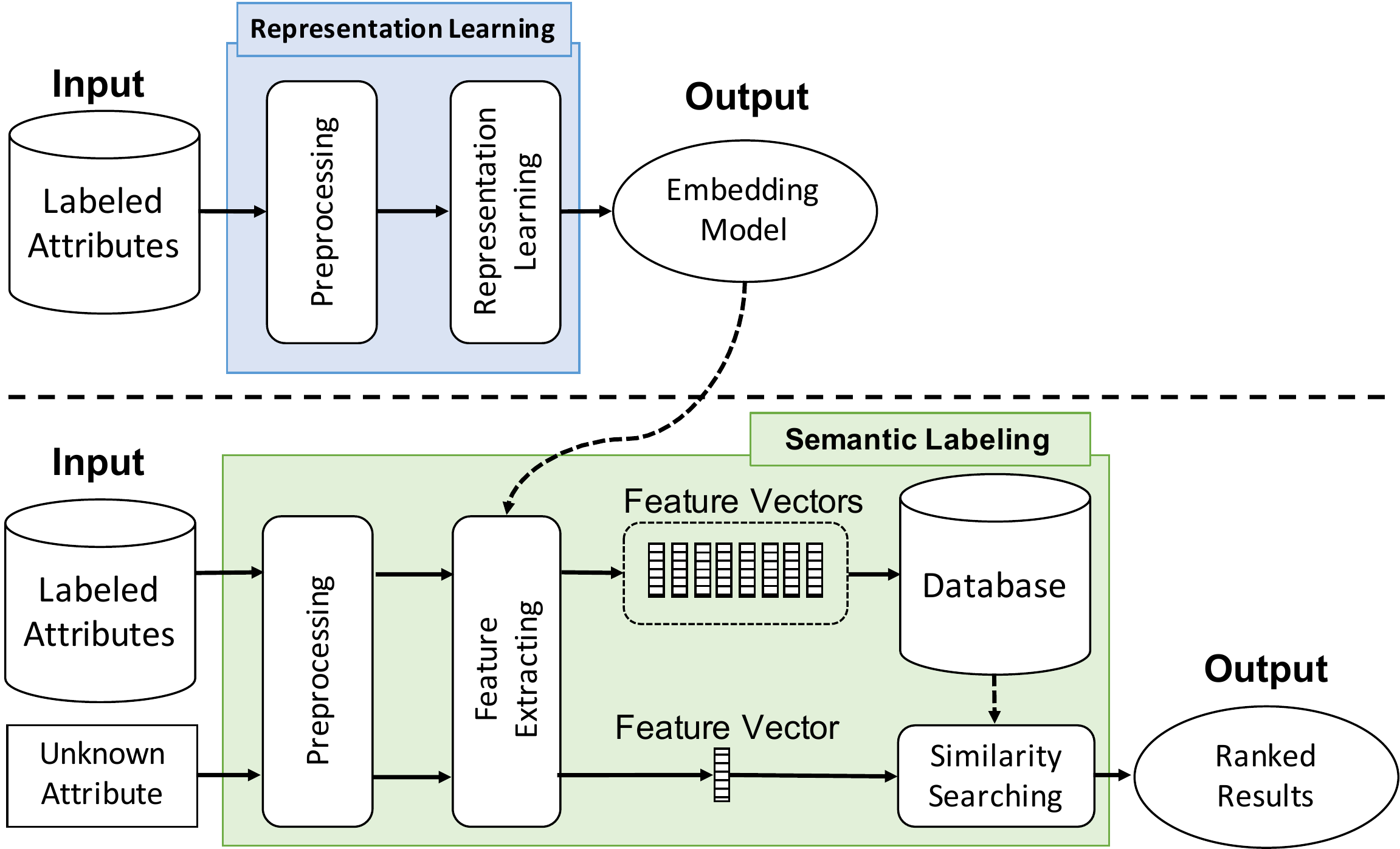}
\caption{The general architecture of semantic labeling with EmbNum}
\label{fig:framework}
\vspace*{-5mm}
\end{figure}

Fig. \ref{fig:framework} depicts the workflow of the semantic labeling task with \textit{EmbNum}which consists of two main phases: \textit{representation learning} and \textit{semantic labeling}. 

In the \textit{representation learning} phase, the \textit{preprocessing} module generate samples from the original distribution of \textit{labeled attributes}. After that, these samples are used as input for the \textit{representation learning} module. The output of the \textit{representation learning} is an \textit{embedding model} which is used in the \textit{feature extracting} module in the \textit{semantic labeling} phase. 

In the \textit{semantic labeling} phase, \textit{labeled attributes} also performed sampling with the \textit{prepossessing} module and then deriving feature vectors with the \textit{feature extracting} module. The embedding model from the previous phase is used as the feature extractor for preprocessed attributes in the feature extracting module. These feature vectors are stored in the feature \textit{database} for future similarity comparison. Suppose we have to perform semantic labeling for an unknown attribute, again the \textit{preprocessing} module and \textit{feature extracting} are performed to get the feature vector.  Next, the \textit{similarity searching} module is performed to calculate the Euclidean distance of the feature vector of the unknown attribute with all the feature vectors in the \textit{database}. Finally, the system returns a ranking list of the most relevant attributes. The label of the top result (the most similar) is assigned to the unknown attribute. 
\subsection{Preprocessing Module} \label{preprocessing}
Because the size of numerical attributes could vary from a few to thousands of values, we use inverse transform sampling \cite{wikiits} to standardize the input size. This technique is chosen because it retains the original distribution of a given list of numerical values. After sampling, the list of numerical values is sorted in a specific order to leverage the capability of the CNN network to learn representations from the cumulative distribution of numerical attributes. The inverse transform sampling is described as follows.
\vspace*{-4mm}
\subsubsection{Inverse Transform Sampling}
Given a numerical attribute $a \in A$ has numerical values $V_a = \{v_1, v_2, v_3, ..., v_n\}$. We treat $V_a$ as a discrete distribution, then the cumulative distribution function (CDF) of $v \in V_a$ is $F_{V_a}(v)$ as follows.
\begin{equation}
F_{V_a}(v) = P(V_a \leqslant v), v \in V_a, F_{V_a}:\mathbb{R} \rightarrow [0,1]
\end{equation}
$P(.)$ represents the probability of values in $V_a$ less than or equal to $v$. The inverse distribution function of $F_{V_a}(.)$ takes the probability $p$ as input and return $v \in V_a$ as follows.
\begin{equation}
    F_{V_a}^{-1}(p) = min\{v: F_{V_a}(v) \geqslant p\}, p \in [0,1]
\end{equation}
\sloppy We select $h$ numbers from $V_a$ with each number is the output of the inverse distribution function $F_{V_a}^{-1}(p)$ with probability $p \in \mathcal{P} = \{ \frac{i}{h}  | i \in \{1, 2, 3, ..., h\}\}$. For example, when $h=100$, then $\mathcal{P} = \{0.01, 0.02, 0.03, ..., 1\}$. For each attribute $a \in A$, we have a preprocessed numerical attribute $x = \{v_1, v_2, v_3, ..., v_h\}$. 

In summary, given a list of attributes $A = \{a_1, a_2, a_3, ..., a_n\}$, after preprocessing, we have a list of preprocessed attributes $X = \{x_1, x_2, x_3,..., x_n\}$. 

\vspace*{-4mm}
\subsubsection{Sampling Analysis} 
In this section, we present an analysis of how well the samples of the inverse transform sampling fit with original data. We also present a comparison of the sampling results of the inverse transform sampling with the random-choice sampling technique, where a random sample was extracted from a given list of numerical values. 

\begin{figure}
\centering
\vspace*{-5mm}
\subfloat[decRainDays]{\label{fig:cdf_a}\includegraphics[width=0.5\textwidth]{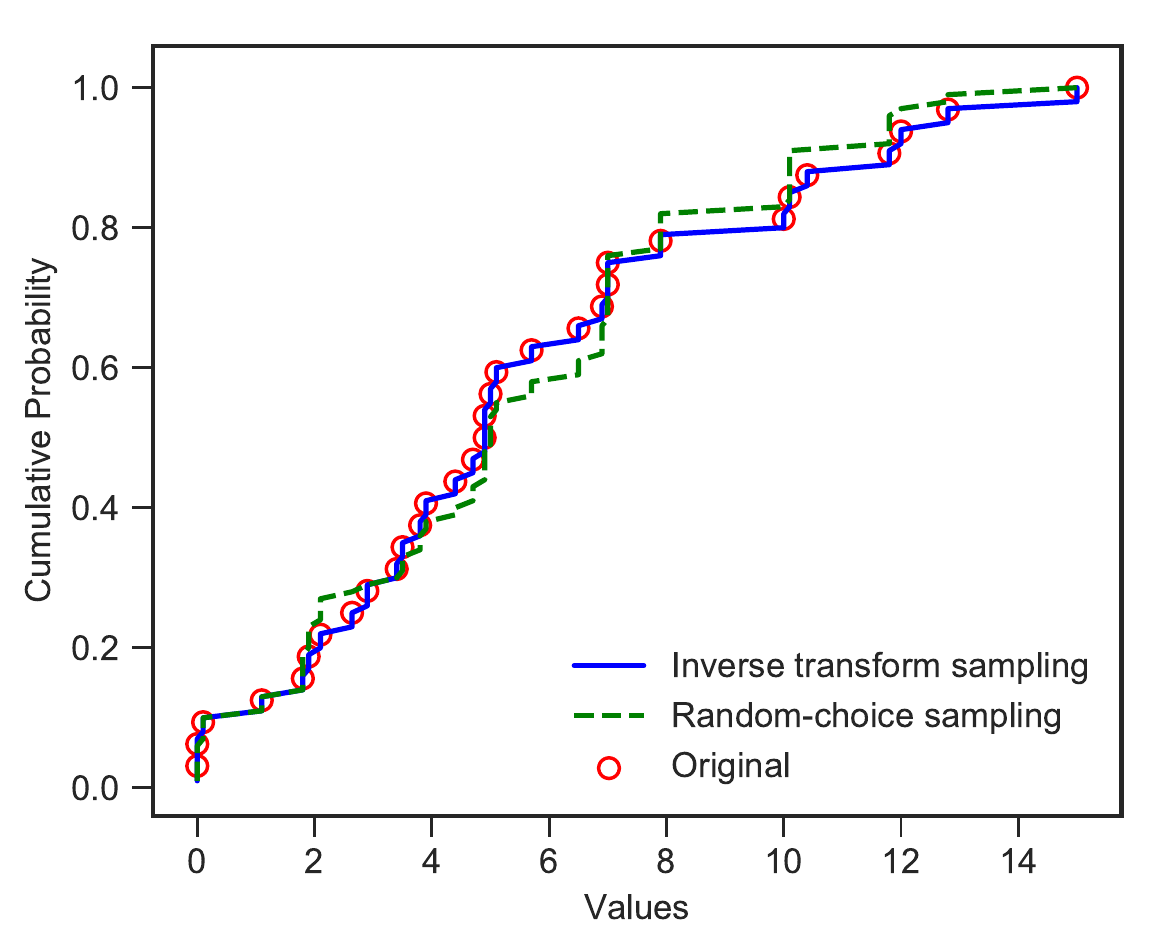}}
\subfloat[aprHighF]{\label{fig:cdf_b}\includegraphics[width=0.5\textwidth]{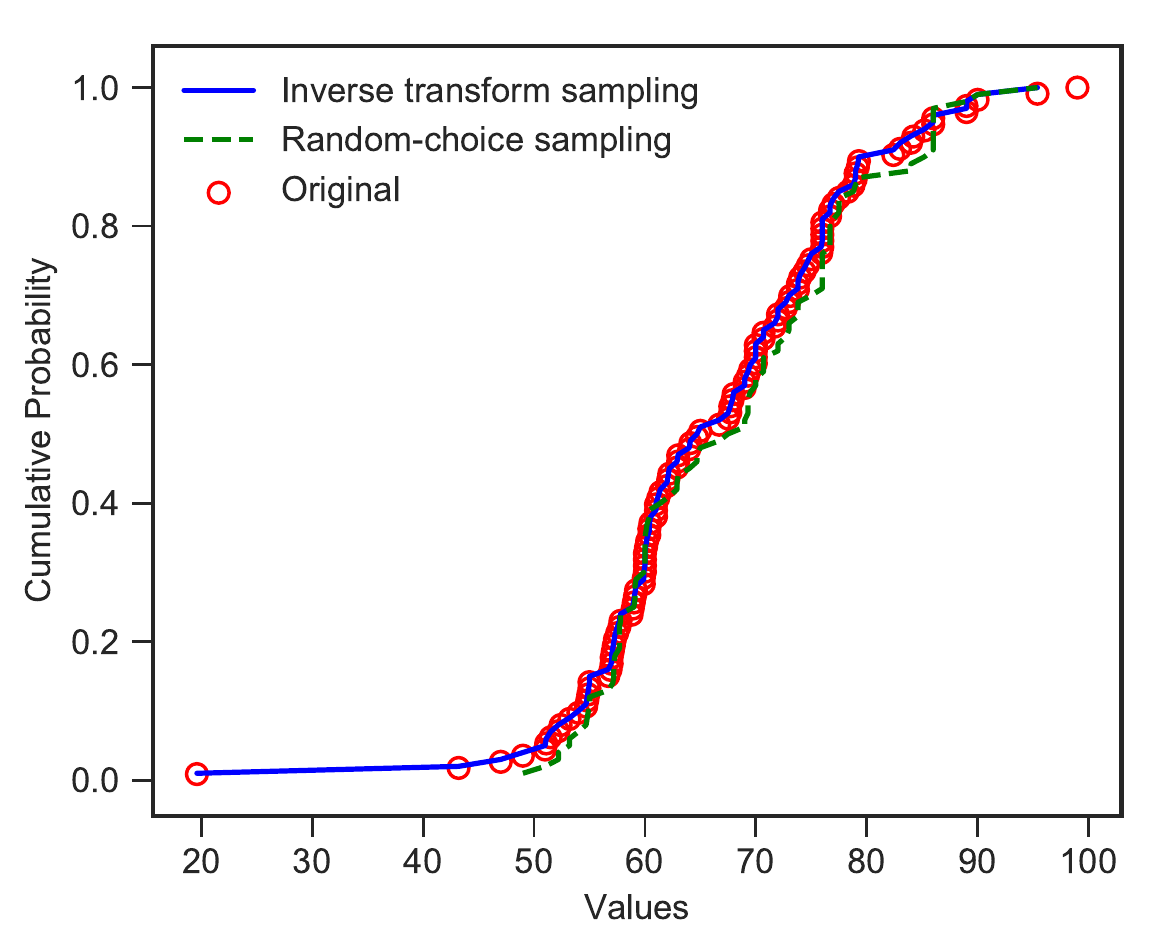}}
\caption{Analysis of inverse transform sampling and random-choice sampling on the \textit{decRainDays} property (\ref{fig:cdf_a}) and the \textit{aprHighF} property (\ref{fig:cdf_b}) of City Data.} 
\label{fig:cdf}
\vspace*{-5mm}
\end{figure}

Fig. \ref{fig:cdf} depicts the sampling results of two sampling techniques on the \textit{decRainDays} property and the \textit{aprHighF} property of City Data. The distribution of sampling from the inverse transforms (the blue curve) clearly fit the original distribution (the red circles) better than the random-choice sampling did (the green curve). Therefore, in our experiment, we used the inverse transform sampling for preprocessing data.

\subsection{Representation Learning Phase} \label{learning}
\begin{figure}
\centering
\vspace*{-5mm}
\includegraphics[width=0.9\textwidth]{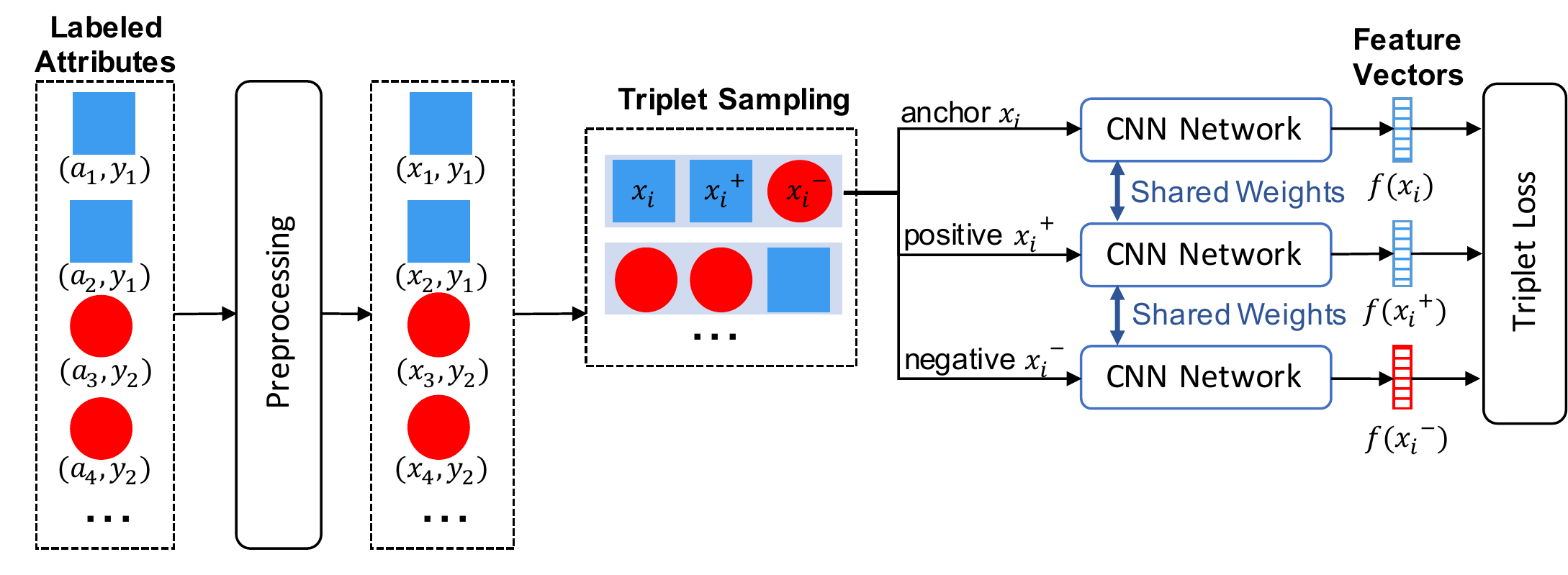}
\caption{Representation learning architecture}
\label{fig:learning_1}
\vspace*{-5mm}
\end{figure}

Fig. \ref{fig:learning_1} depicts the architecture of representation learning phase given a list vectors $X = \{ x_1, x_2, x_3, ..., x_n\}$ of $n$ numerical attributes and their $m$ semantic labels $Y = \{ y_1, y_2, y_3, ..., y_m\}$. The blue square, and red circle indicate the semantic label $y_1$ and the semantic label $y_2$, respectively, where $y_1 \neq y_2$, .

We used a triplet network \cite{schroff2015facenet} to learn a $f(.)$ function to map a numerical attribute into an embedding space. For an input $x$, $f(x)$ is the output of a representation learning network to convert $x$ into a $k$ dimensions Euclidean space, $f(x) \in \mathbb{R}^k$. The similarity distance between two numerical attributes $x_i$ and $x_j$ is calculated by using the Euclidean distance between $f(x_i)$ and $f(x_j)$:
\begin{equation}
    d_f(x_i,x_j) = d(f(x_i),f(x_j)) = ||f(x_i) - f(x_j)||_2^2 = \sqrt{\sum_{k=1}^m (f(x_i)_k - f(x_j)_k)^2}
\end{equation}
A triplet $(x, x^+, x^-)$ with $(x,x^+, x^- \in X)$ is a combination of a numerical attribute $x$ where the semantic label is $y$, a similar attribute $x^+$ where the semantic label is $y$, and a dissimilar attribute $x^-$ where the semantic label is not $y$. The key idea of a triplet network relies on the empirical observation that the distance between the positive pair must be less than the distance between the negative pair $d_f(x,x^+) < d_f(x,x^-)$ \cite{schroff2015facenet}. Then, the loss function for the triplet network is defined as follows.
\begin{equation}
L = max(0, \alpha + d_f(x,x^+) - d_f(x,x^-))
\end{equation}
where $\alpha$ is a hyperparameter that regularizes between positive pair distance and negative pair distance. 

We utilized the hard negative sampling method \cite{schroff2015facenet} to select triplets for training. The hard negative sampling is a technique of choosing the closest sample with an anchor among the dissimilar attributes in a mini-batch of learning. It will help the training process to converge faster. 

\vspace*{-4mm}
\subsubsection{CNN Network Architecture:}
Many CNN architectures have been designed to capture latent features directly from data. In this work, we used ResNet 18 \cite{he2016deep} because it provides good accuracy and requires fewer parameters to train \cite{canziani2016analysis}. Its architecture used ReLU as a non-linear activation function. To normalize the distribution of each input features in each layer, we have also used batch normalization \cite{ioffe2015batch} after each convolution, before each ReLU activation function. Because input data were one-dimension, we modified the structure of convolutional to one-dimension on convolutional layers, batch normalization layers, and pooling layers. The output of the network is a vector with $k$  dimensions. 

\subsection{Semantic Labeling Phase} \label{labeling}
\begin{figure}
\centering
\vspace*{-5mm}
\includegraphics[width=0.9\textwidth]{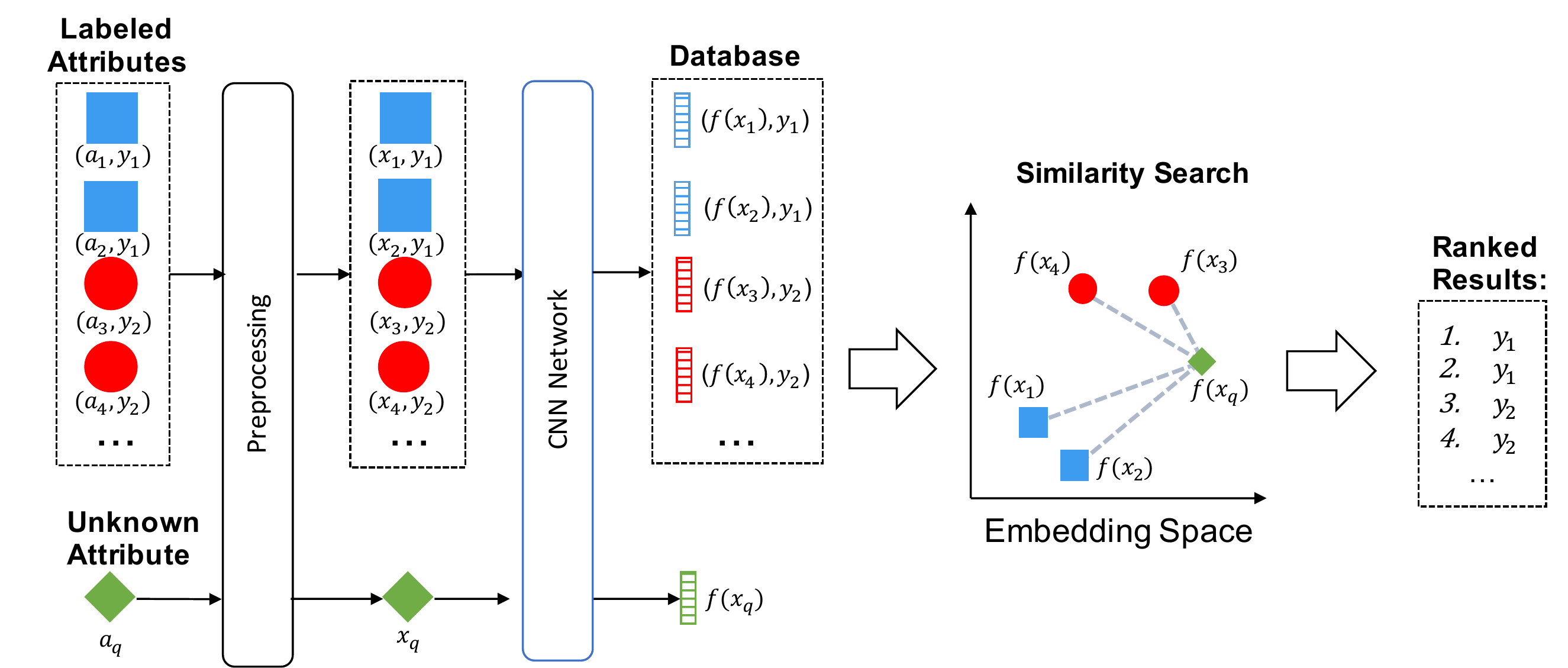}
\caption{Example of semantic labeling for an unknown attribute}
\label{fig:labeling_1}
\vspace*{-5mm}
\end{figure}

Fig. \ref{fig:labeling_1} depicts an example of semantic labeling for an unknown attribute. Given labeled attributes and their semantic labels
$\{(a_1,y_1), (a_2,y_2), (a_3,y_3),..., (a_n, y_m)\}$, first, the data were preprocessed as $\{(x_1,y_1), (x_2,y_2), (x_3,y_3),..., (x_n, y_m)\}$. After that, we used the mapping function $f(.)$ to map all the labeled data to embedding space $\{(f(x_1),y_1), (f(x_2),y_2), (f(x_3),y_3),..., (f(x_n),y_m)\}$. Those data were stored in the database $D$ for the next comparison in the semantic labeling process. 

Given an unknown attribute $a_{q}$, we performed the preprocessing to get $x_q$ and the feature extracting to get $f(x_q)$. After that, we compared the Euclidean distance from $f(x_q)$ to each of the labeled data in $D$. Then we get a ranking list of the most similar attributes with $f(x_q)$. We assigned the semantic label of the top one of the ranking result to the unknown data.
\section{Evaluation} \label{experiment}
In this section, we first describe benchmark datasets in section \ref{data} and evaluation metrics in Section \ref{metric}. Next, the details about implementation of representation learning as well as the visualization of embedding features for numerical attributes are presented in Section \ref{res:learning}. Finally, we report the experimental results of semantic labeling in terms of effectiveness and efficiency in Section \ref{res:labeling}. 
\subsection{Benchmark Datasets} \label{data}
To evaluate our proposed approach, we used two datasets i.e., City Data and Open Data. City Data is the standard data used in the previous studies \cite{pham16:iswc}, \cite{ramnandan2015assigning}. Open Data is a newly built dataset extracted from tables of Open Data portals. All the table data were evaluated on the task of numerical attribute semantic labeling. The detailed statistics of each dataset are shown in Table \ref{tab:data}
\begin{table}
\vspace*{-5mm}
\centering
\caption{Description of City Data and Open Data}
\label{tab:data}
\begin{tabular}{@{}l|c|c|c|crrr@{}}
\toprule
\multirow{2}{*}{} & \multirow{2}{*}{\begin{tabular}[c]{@{}c@{}}\# \\ Sources\end{tabular}} & \multirow{2}{*}{\begin{tabular}[c]{@{}c@{}}\# \\ Labels\end{tabular}} & \multirow{2}{*}{\begin{tabular}[c]{@{}c@{}}\# \\ Attributes\end{tabular}} & \multicolumn{4}{c}{\# Rows of Each Attribute}                                                              \\
                  &                             &                                                                                &                                & Min                   & \multicolumn{1}{c}{Max} & \multicolumn{1}{c}{Median} & \multicolumn{1}{c}{Average} \\ \midrule
City Data         & \multicolumn{1}{c|}{10}     & \multicolumn{1}{c|}{30}                                                        & \multicolumn{1}{c|}{300}       & \multicolumn{1}{r}{4} & 2,251                    & 113                        & 642.73                      \\
Open Data         & \multicolumn{1}{c|}{10}     & \multicolumn{1}{c|}{50}                                                        & \multicolumn{1}{c|}{500}       & \multicolumn{1}{r}{4} & 186,082                  & 467                        & 14,659.63                    \\ \bottomrule
\end{tabular}
\vspace*{-5mm}
\end{table}

\vspace*{-4mm}
\subsubsection{The City Data} \cite{ramnandan2015assigning} has 30 numerical properties extracted from the city class in DBpedia. The dataset consists of 10 sources; each source has 30 numerical attributes associated with 30 data properties. 
\vspace*{-4mm}
\subsubsection{The Open Data} has 50 numerical properties extracted from the tables in five Open Data Portals. We built the dataset to test semantic labeling for numerical values in the open environment.

To build the dataset, we extracted table data from five Open Data portals, i.e., Ireland (data.gov.ie), the UK (data.gov.uk), the EU (data.europa.eu), Canada (open.canada.ca), and Australia (data.gov.au). First, we crawled CSV files from the five Open Data portals and selected files that had their sizes less than 50 MB. Then, we analyzed tables in CSV files and selected only numerical attributes. After that, we created categories of numerical attributes based on the clustering of the numerical attributes with respect to the textual similarity of column headers. We got 7,496 categories in total. 

We manually evaluated these categories with two criteria: (1) The first criterion was to pick up categories with a certain frequency. By examining the collection of data, we found that high-frequency categories are often unclear on their semantics, while low-frequency categories are often unstable as data. We decided to pick up categories with ten frequency by following the setting of City Data. (2) The second criterion was removing the categories where column headers had too general meanings such as ``ID," ``name," or ``value."

Finally, we chose 50 categories as semantic labels; each semantic label had ten numerical attributes. Following the guideline of City Data, we also made 10 data sources by combining each numerical attributes from each category. The dataset is available at https://github.com/phucty/embnum. 

\subsection{Evaluation Metrics} \label{metric}
We used the mean reciprocal rank score (MRR) to measure the effectiveness of semantic labeling. The MRR score was used in the previous studies \cite{ramnandan2015assigning}, \cite{pham16:iswc} to measure the probability correctness of a ranking result list. To measure the efficiency of \textit{EmbNum}over the baseline methods, we evaluated the run-time in seconds of the semantic labeling process.

\subsection{Representation Learning Evaluation}\label{res:learning}
We randomly divided City Data into two equal parts as 50\% for learning a similarity metric, and 50\% for evaluating the task of semantic labeling. The first part was used representation learning of EmbNum, and it had also used to learn the similarity metric for DSL. To learn the similarity metric for DSL, we followed their guideline that using logistic regression to train the similarity metrics with training samples are the pairs of numerical attributes. 

We used PyTorch (http://pytorch.org) to implement the framework and experiments. We trained the network using stochastic gradient descent (SGD) with back-propagation, a momentum of 0.9, and a weight decay of 1E-5. We started with a learning rate of 0.01 and reduced it with a step size of 10 to finalize the model. We set the dimension of the attribute input vector $h$ and the attribute output vector $k$ as 100. Notice that we did not perform finding the optimal hyper-parameters of $h$, and $k$, the optimal hyper-parameter searching are left for future work. The representation learning was trained with 100 epochs. After each epoch, we evaluated the task of semantic labeling on the MRR score using the training data. We saved the learned model having the highest MRR score. 

All of our experiments ran on Dell Precision 7710 with an Intel Xeon E3-1535M-CPU, 16 GB of RAM, and an NVIDIA Quadro M3000M GPU. The training time of \textit{EmbNum} is 3,486 seconds while training time of DSL is 196 seconds. It is clear that \textit{EmbNum} used with the deep learning approach needed more times to train the similarity metric than logistic regression with the DSL approach. However, the similarity metric is only needed to train once, and it could be applied to other domains without retraining.

\subsubsection{Visualization of Embedding Features}
\begin{figure}
\centering
\vspace*{-5mm}
\includegraphics[width=\textwidth]{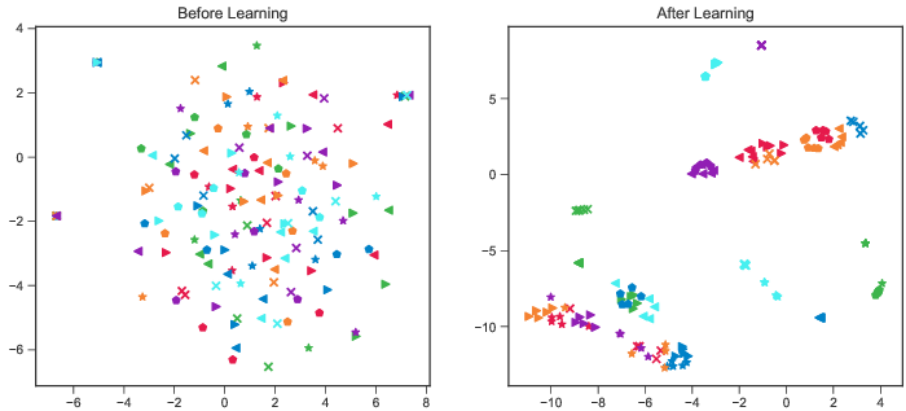}
\caption{t-SNE visualization \cite{maaten2008visualizing} of City Data embedding vectors} 
\label{fig:emb}
\vspace*{-5mm}
\end{figure}

Fig. \ref{fig:emb} shows the visualization of the embedding features of numerical attributes from City Data. We used t-SNE visualization \cite{maaten2008visualizing} that projects the output features of the representation model from $k$ dimensions into two dimensions. The left figure depicts the state before learning the representation, while the right figure depicts the state after learning. It is clear from the visualization that the distance between two similar semantic labels is close, and the distance between different semantic labels is far after the representation learning. It means the learning process has learned good representations for numerical attributes.
\subsection{Semantic Labeling Evaluation} \label{res:labeling}
This section describes the evaluation of semantic labeling in terms of effectiveness and efficiency of \textit{EmbNum} and the two baseline approaches: SemanticTyper \cite{ramnandan2015assigning} and DSL \cite{pham16:iswc}. 

\subsubsection{Experimental Setting} Suppose a dataset $S = \{s_1, s_2, s_3, ..., s_d\}$ has $d$ data sources. One data source was retained as the unknown data, and the remaining $d-1$ data sources were used as the labeled data. We repeated this process $d$ times, with each of the data source used exactly once as the unknown data.
Additionally, we set the number of sources in the labeled data increase from one source to $d-1$ sources to analyze the effect of increment in the number of labeled data on the performance of semantic labeling. It is noticed that we tested all possible combinations of labeled sources. We obtained the MRR scores and labeling times on $d \times (2^{d-1}-1)$ experiments and then averaged them to produce the $d-1$ estimations of the number of sources in the labeled data.

Table \ref{tab:setting} depicts the experiment setting on City Data with five data sources. From $1^{st}$ experiment to $15^{th}$ experiment, $s_1$ is assigned as the unknown sources, the remaining sources are the labeled sources. The labeled data which have one source is tested in the $1^{st}\text{-}4^{th}$ experiment, whereas in the $5^{th}\text{-}15^{th}$, more labeled sources are tested. We conducted a similar approach for the remaining experiments. Overall, we performed 75 experiments on the five sources of City Data, and 5,110 experiments on the ten sources of Open Data. 

\begin{table}
\centering
\caption{Experiment setting of semantic labeling on City Data with five data sources}
\label{tab:setting}
\begin{tabularx}{\textwidth}{|l|Y|Y|Y|Y|Y|Y|Y|Y|Y|Y|Y|Y|Y|Y|Y|Y|Y|Y|Y}
\hline
Experiment     & 1 & 2 & 3 & 4 & 5 & 6 & 7 & 8 & 9 & 10 & 11 & 12   & 13 & 14 & 15 & 16 & ... & 75\\ \hline
\begin{tabular}[l]{@{}l@{}}Unknown\\Source\end{tabular} & $s_1$ & $s_1$ & $s_1$ & $s_1$ & $s_1$ & $s_1$ & $s_1$ & $s_1$ & $s_1$ & $s_1$ & $s_1$ & $s_1$ & $s_1$ & $s_1$ & $s_1$ & $s_2$ & ... & $s_5$ \\ \hline
\multirow{4}{*}{\begin{tabular}[l]{@{}c@{}}Labeled\\ Sources\end{tabular}} & $s_2$ & $s_3$ & $s_4$ & $s_5$ & $s_2$ & $s_2$ & $s_2$ & $s_3$ & $s_3$ & $s_4$ & $s_2$ & $s_2$ & $s_2$ & $s_3$ & $s_2$ & $s_1$ & ... & $s_1$  \\
&  &  &  &  & $s_3$ & $s_4$ & $s_5$ & $s_4$ & $s_5$ & $s_5$ & $s_3$ & $s_3$ & $s_4$ & $s_4$ & $s_3$ & & ... & $s_2$ \\
& & & & & & & & & & & $s_4$ & $s_5$ & $s_5$ & $s_5$ & $s_4$ & & ...& $s_3$  \\
& & & & & & & & & & & & & & & $s_5$ & & ... & $s_4$  \\ \hline
\end{tabularx}
\vspace*{-5mm}
\end{table}

\subsubsection{Experimental Results Regarding Effectiveness}
\begin{figure}
\centering
\vspace*{-5mm}
\subfloat[City Data]{\label{fig:acc_a}\includegraphics[width=0.5\textwidth]{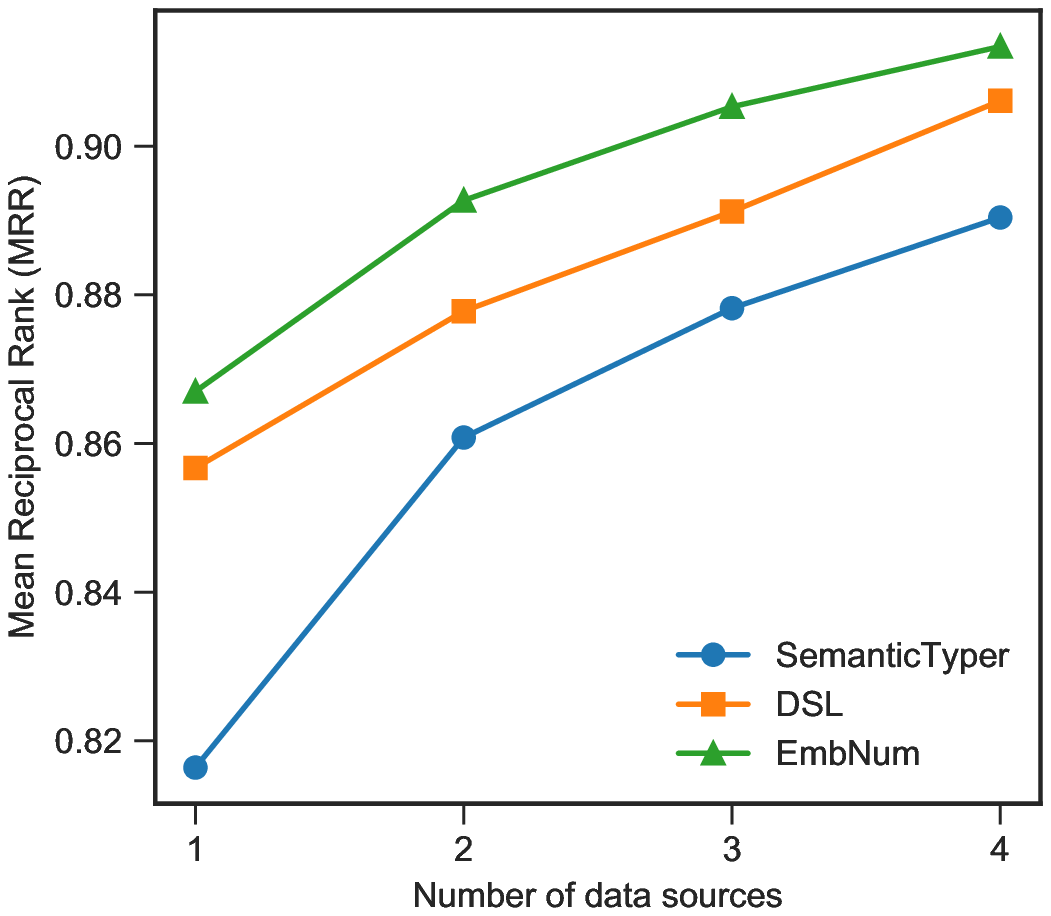}}
\subfloat[Open Data]{\label{fig:acc_b}\includegraphics[width=0.5\textwidth]{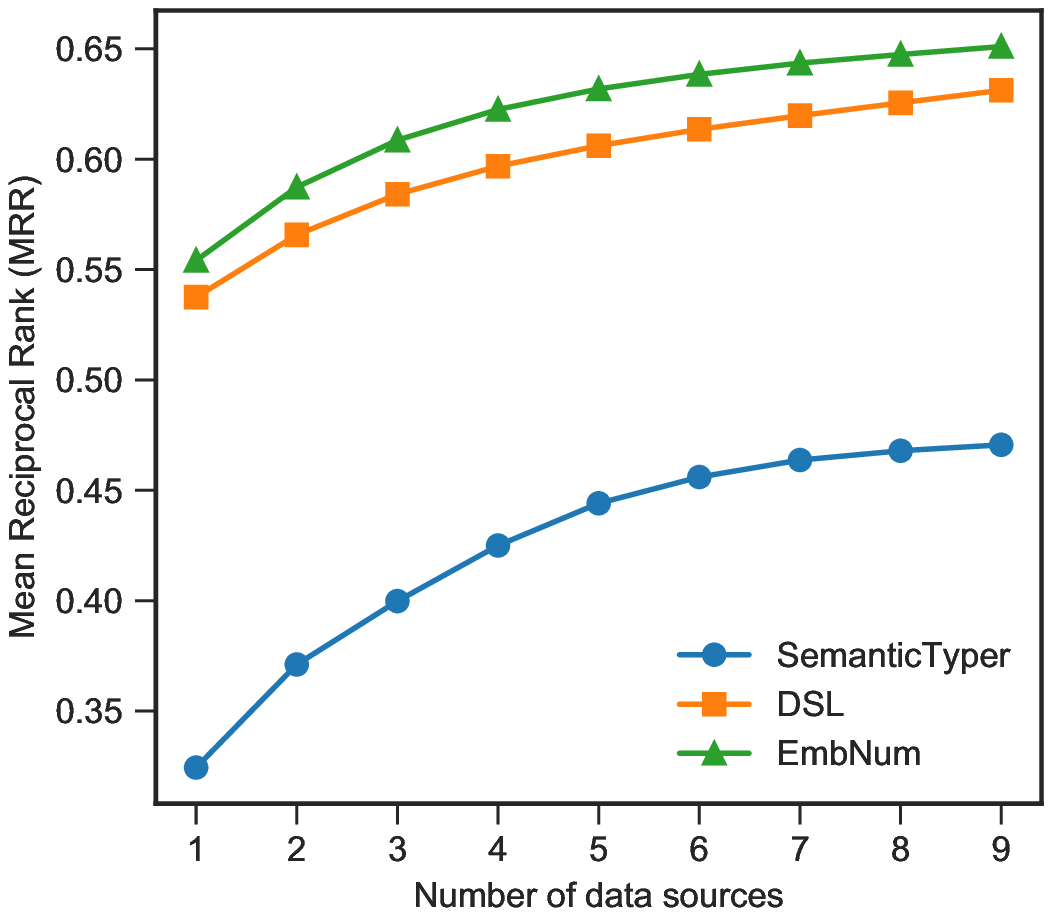}}
\caption{Semantic Labeling in the MRR score on City Data and Open Data}
\label{fig:acc}
\vspace*{-5mm}
\end{figure}

We tested SemanticTyper, DSL, and \textit{EmbNum} on the semantic labeling task using the MRR score to evaluate the effectiveness. The results are shown in Table \ref{tab:mrr} and Fig. \ref{fig:acc}. 
\begin{table}
\centering
\vspace*{-5mm}
\caption{Semantic Labeling in the MRR score on City Data and Open Data}
\label{tab:mrr}
\begin{tabular}{c|l|ccccccccc}
\multirow{2}{*}{Data}      & \multicolumn{1}{c|}{\multirow{2}{*}{Method}} & \multicolumn{9}{c}{Labeled Sources}                                                                                                                    \\
                           & \multicolumn{1}{c|}{}                        & 1              & 2              & 3              & 4              & 5              & 6              & 7              & 8              & 9              \\ \hline
\multirow{3}{*}{City} & SemanticTyper\cite{ramnandan2015assigning}                                & 0.816          & 0.861          & 0.878          & 0.890          & -              & -              & -              & -              & -              \\
                           & DSL\cite{pham16:iswc}                                          & 0.857          & 0.861          & 0.8978          & 0.906          & -              & -              & -              & -              & -              \\
                           & EmbNum                                       & \textbf{0.867} & \textbf{0.893} & \textbf{0.905} & \textbf{0.913} & -              & -              & -              & -              & -              \\ \hline
\multirow{3}{*}{Open} & SemanticTyper\cite{ramnandan2015assigning}                                          & 0.324          & 0.371          & 0.4070          & 0.425          & 0.444          & 0.456          & 0.464          & 0.468          & 0.471          \\
                           & DSL\cite{pham16:iswc}                                          & 0.538          & 0.566          & 0.584          & 0.597          & 0.606          & 0.614          & 0.620          & 0.626          & 0.631          \\
                           & EmbNum                                       & \textbf{0.554} & \textbf{0.587} & \textbf{0.609} & \textbf{0.623} & \textbf{0.632} & \textbf{0.638} & \textbf{0.644} & \textbf{0.647} & \textbf{0.651}
\end{tabular}
\vspace*{-5mm}
\end{table}

The MRR scores obtained by three methods steadily increase along with the number of labeled sources. It suggests that the more labeled sources in the database, the more accurate the assigned semantic labels are. DSL outperformed SemanticTyper because DSL uses a combined metric that includes multiple similarity features, while SemanticTyper uses only the KS test. \textit{EmbNum} learned directly from the input without making any assumption on data type and data distribution, hence, outperformed SemanticTyper and DSL on both of datasets. The similarity metric based on a specific hypothesis test, which was used in SemanticTyper and DSL, is not optimized for semantic meanings with various data types and distributions.

To understand whether \textit{EmbNum} does significantly outperform SemanticTyper and DSL, we performed a paired sample t-test on the experiments of City Data and Open Data. We set the cutoff value for determining statistical significance to 0.01. The results of the paired t-test revealed that \textit{EmbNum} significantly outperforms SemanticTyper $(p<0.0001)$ and DSL $(p=0.0068)$ on City Data. The similar results could be obtained on Open Data, where \textit{EmbNum} significantly outperforms SemanticTyper $(p<0.0001)$ and DSL $(p<0.0001)$.

\subsubsection{Experimental Results Regarding Efficiency} 
\begin{figure}
\centering
\vspace*{-5mm}
\subfloat[City Data]{\label{fig:s_a}\includegraphics[width=0.5\textwidth]{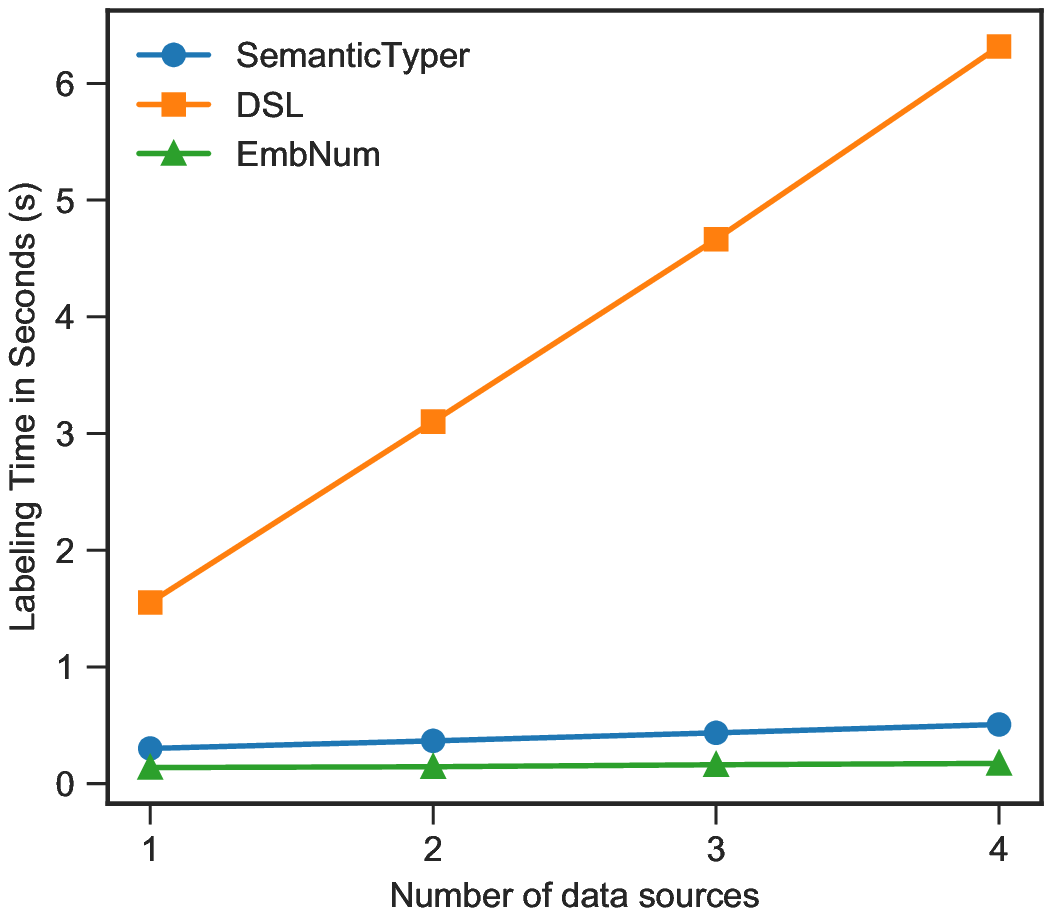}}
\subfloat[Open Data]{\label{fig:s_b}\includegraphics[width=0.5\textwidth]{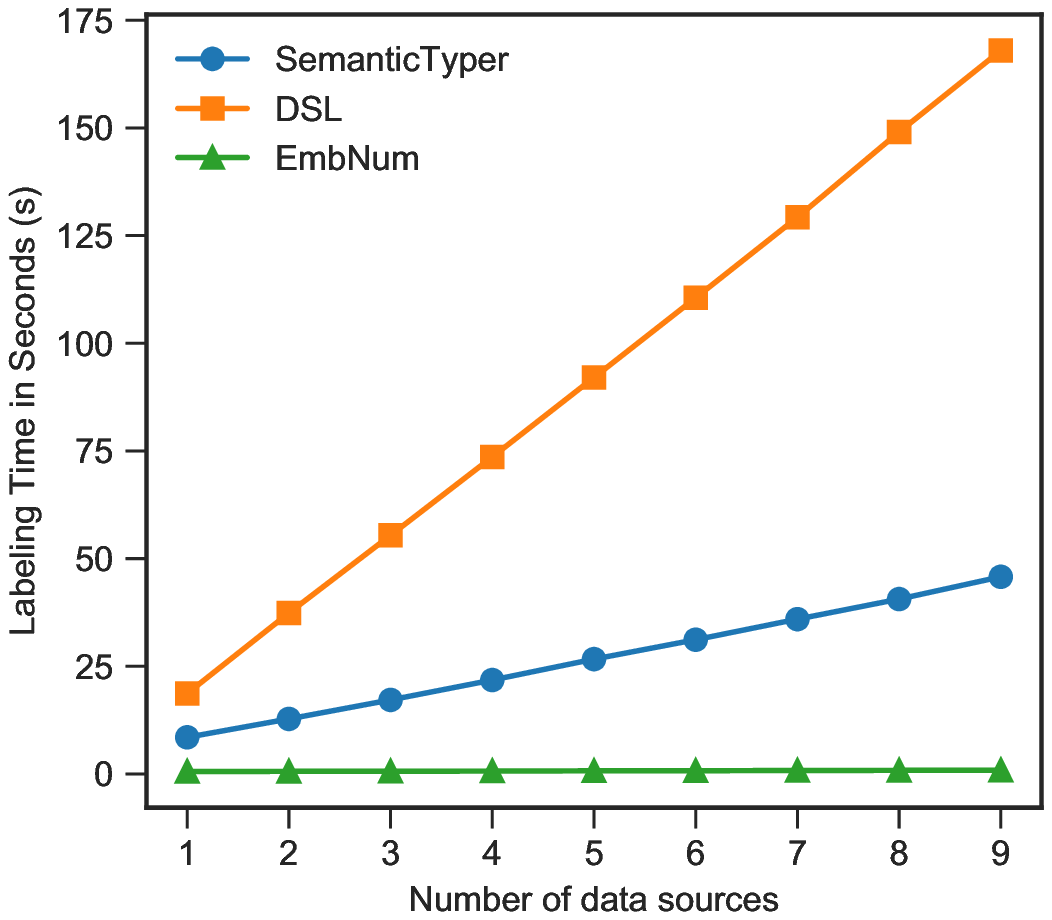}}
\caption{Run-time in seconds of semantic labeling on City Data and Open Data} 
\label{fig:s}
\vspace*{-5mm}
\end{figure}

Table \ref{tab:s} and Fig. \ref{fig:s} depict the run-time of semantic labeling on SemanticTyper, DSL, and EmbNum. 
\begin{table}
\vspace*{-5mm}
\centering
\caption{Run-time in seconds of semantic labeling on City Data and Open Data}
\label{tab:s}
\begin{tabular}{l|l|rrrrrrrrr}
\multicolumn{1}{c|}{\multirow{2}{*}{Data}} & \multicolumn{1}{c|}{\multirow{2}{*}{Method}} & \multicolumn{9}{c}{Labeled Sources}                                                                                                                    \\
\multicolumn{1}{c|}{}                      & \multicolumn{1}{c|}{}                        & 1              & 2              & 3              & 4              & 5              & 6              & 7              & 8              & 9              \\ \hline
\multirow{3}{*}{City}                 & SemanticTyper\cite{ramnandan2015assigning}                                & 0.3          & 0.4          & 0.4          & 0.5          & -              & -              & -              & -              & -              \\
                                           & DSL\cite{pham16:iswc}                                          & 1.6          & 3.1          & 4.7          & 6.3          & -              & -              & -              & -              & -              \\
                                           & EmbNum                                       & \textbf{0.1} & \textbf{0.1} & \textbf{0.2} & \textbf{0.2} & -              & -              & -              & -              & -              \\ \hline
\multirow{3}{*}{Open}                 & SemanticTyper\cite{ramnandan2015assigning}                                & 8.5          & 12.8         & 17.2         & 21.8         & 26.7         & 31.2         & 35.9         & 40.6         & 45.8         \\
                                           & DSL\cite{pham16:iswc}                                          & 18.7         & 37.4         & 55.4         & 73.6         & 92.1         & 110.6        & 129.2        & 149.1        & 168        \\
                                           & EmbNum                                       & \textbf{0.6} & \textbf{0.6} & \textbf{0.6} & \textbf{0.7} & \textbf{0.7} & \textbf{0.8} & \textbf{0.8} & \textbf{0.8} & \textbf{0.9}
\end{tabular}
\vspace*{-5mm}
\end{table}

The run-time of semantic labeling linearly increases with the number of labeled sources. The run-time of DSL was extremely high when the number of labeled data sources increased because three similarity metrics were needed to be performed. The run-time of SemanticTyper was less than DSL because it only used the KS test as a similarity metric. Semantic labeling with \textit{EmbNum} is significantly faster than SemanticTyper (about 25 times), and DSL (about 92 times). \textit{EmbNum} outperforms the baseline approaches in run-time since the similarity metric of \textit{EmbNum} was calculated directly on extracted feature vectors instead of all original values. 

\section{Related Work} \label{relate_work}
In this section, we present the previous approaches for semantic labeling with textual information and numerical information. Next, we will briefly present related works on representation learning.

\paragraph{Semantic labeling with textual information:}
These studies address the problem of semantic labeling for table attributes using the information on header labels and textual values \cite{Ermilov2013, Zhang2013, ritze2015matching}. The most common technique used entity linkage for mapping textual values of attributes to entities in a knowledge base. After that, the schema of entities was used to find the semantic concept for table attributes. Also, the additional textual descriptions of tables were considered in the studies of \cite{Adelfio2013, Venetis2011, Wang2012} which improve the performance of the labeling task. Beside textual information,  numerical information should be used in other ways to build an effective integrated system.

\paragraph{Semantic labeling with numerical information:} Several attempts have been made on using descriptive statistics with hypothesis tests as similarity metrics to compare the similarity of numerical attributes. Stonebraker et al. \cite{Stonebraker2013} used the Welch’s t-test \cite{lehmann2006testing} as a similar metric to compare attributes of numerical data. SemanticTyper \cite{ramnandan2015assigning} used the Kolmogorov Skmiro test (KS test) \cite{lehmann2006testing} to compare the empirical distribution of numerical attributes. SemanticTyper achieved better performance than using the Welch's t-test. Minh Pham et al. \cite{pham16:iswc} (DSL) extended SemanticTyper by proposing a new similarity metric that is a combination of the KS test and two other metrics: the Mann-Whitney test (MW test) and the numeric Jaccard similarity. Their experiments showed that using the combined metric provided better results over using only a KS test. Neumaier et al. \cite{neumaier2016multi} created a numeric background knowledge base from DBpedia. Given an unknown numerical attribute, they used the KS test as a similarity metric to compare with each labeled attribute in the numeric background knowledge base. Overall, the similarity metrics used in these approaches are hypothesis tests which are calculated under a specific assumption about data type, and data distribution. In contrast to these approaches, we propose a neural numerical embedding model to learn a similarity metric directly from data without making any assumption about data type and data distribution. 

\paragraph{Representation learning:}
Deep metric learning has achieved considerable success in extracting useful representations of data \cite{Bengio-2009, zeiler2014visualizing, sermanet2013overfeat}. Moreover, the features extracted with deep neural networks can be used for other tasks \cite{sharif2014cnn}. Florian et al. proposed a triplet network that takes an anchor, a positive sample (of the same class as the anchor), and negative sample (of a different class than an anchor) examples, to learn a mapping function to embed the anchor closer to the positive example than the negative example \cite{schroff2015facenet}. Herman et al. \cite{hermans2017defense} proposed a method for triplet mining that selects hard examples for the learning network. It will help in training the network more efficiently. In this study, we used a representation network, which is a combination of a CNN network and triplet network, to learn a metric to measure the similarity between numerical attributes. We also introduced a sampling technique to handle the issue of varying input size of numerical attributes. 

\section{Conclusion}\label{conclusion}
In this paper, we proposed the \textit{EmbNum} model to learn a useful representation and a similarity metric for numerical attributes. We also proposed a preprocessing method by sampling the numerical values in attributes. The experimental results of the representation learning show that the embedding model can capture good latent features from numerical attributes. In a task of semantic labeling, the performance results of \textit{EmbNum} significantly outperforms the baseline approaches on City Data, and Open Data. Additionally, our model does not assume distribution forms of data while the baseline approaches do. As a result, \textit{EmbNum}is more suitable to apply open data where we do not know data distribution forms in advance. 

In future work, we plan to extend our work in the three directions. The first direction is to extend the similarity metric to interpret multiple scales. In this study, we assume that two numerical attributes are similar when they are expressed on the same scale. In fact, two attributes in the same meaning could be measured using different scales. For instance, the numerical attributes ``human height" could be expressed in ``centimeters," but they could also be expressed in ``feet." The second direction is to extend the metric to interpret the hierarchical representation of numerical data. The current presentation using the Euclidean distance cannot reflect the hierarchical structure. Building a similarity metric that is hierarchy-aware can help to make a more fine-grained semantic labeling. The third direction is to detect unseen semantic labels that are not available in labeled data. Although we can use \textit{EmbNum} as a similarity metric to cluster unseen attributes before performing the task of semantic labeling, it cannot recognize new semantic types in the labeled data. In \textit{EmbNum}, we assign the top one semantic types of labeled data for an unknown attribute. If the unknown attribute is a new semantic type, it will return an incorrect result. In future work, we will identify the suitable threshold for the similarity metric to recognize new semantic types. 
\bibliographystyle{splncs04}
\bibliography{samplepaper.bib}
\end{document}